\documentclass{jfm}

\usepackage{graphicx}
\usepackage{natbib}
\usepackage{amsmath,amsfonts}
\usepackage{url}
\newcommand\bp{{\mathbf p}}              
\newcommand\bq{{\mathbf q}}              
\newcommand\bx{{\mathbf x}}              
\newcommand\bu{{\mathbf u}}              
\newcommand\bnabla{{\mathbf \nabla}}     
\newcommand\pnabla{\bnabla_\bp}          
\newcommand\bOmega{{\boldsymbol \Omega}}
\newcommand\bGamma{{\mathbf E}}

\title[Exact tensor closures]{Exact tensor closures for the three dimensional Jeffery's equation}
\author[S. Montgomery-Smith, W. He, D.A. Jack, and D.E. Smith]
{B\ls y\ns S\ls T\ls E\ls P\ls H\ls E\ls N\ns M\ls O\ls N\ls T\ls G\ls O\ls M\ls E\ls R\ls Y\ls -\ls S\ls M\ls I\ls T\ls H$^1$\ls,
W\ls E\ls I\ns H\ls E$^1$\ls,
D\ls A\ls V\ls I\ls D\ns A.\ns J\ls A\ls C\ls K$^2$\ls \and
D\ls O\ls U\ls G\ls L\ls A\ls S\ns E.\ns S\ls M\ls I\ls T\ls H$^3$\ns}
\affiliation{$^1$Department of Mathematics, University of Missouri, Columbia MO 65211, U.S.A. \\ [\affilskip]
             $^2$Department of Mechanical Engineering, Baylor University, Waco, TX 76798, U.S.A. \\ [\affilskip]
             $^3$Department of Mechanical and Aerospace Engineering, University of Missouri, Columbia MO 65211, U.S.A.}
\date{30 June 2010}
\volume{000}
\pagerange{\pageref{firstpage}--\pageref{lastpage}}
\begin{document}
\label{firstpage}
\maketitle
\begin{abstract}
This paper presents an exact formula for calculating the fourth-moment tensor from the second-moment tensor for the three dimensional Jeffery's equation. Although this approach falls within the category of a moment tensor closure, it does not rely upon an approximation, either analytic or curve fit, of the fourth-moment tensor as do previous closures. This closure is orthotropic in the sense of \cite{cintra:95}, or equivalently, a natural closure in the sense of \cite{verleye:93}.  The existence of these explicit formulae has been asserted previously, but as far as the authors know, the explicit forms have yet to be published.  The formulae involve elliptic integrals, and are valid whenever fiber orientation was isotropic at some point in time.  Finally, this paper presents the \emph{Fast Exact Closure} (FEC), a fast and in principle exact method for solving Jeffery's equation, which does not require approximate closures, nor the elliptic integral computation.
\end{abstract}



\section{Introduction}
\label{Introduction_Section}
Jeffery's equation \cite[][]{jeffery:23} concerns the motion of the direction of an axi-symmetric rigid body (henceforth referred to as a fiber) under the influence of a Stokesian fluid whose velocity field is $\bu = \bu(\bx,t)$, and under the assumption that the size of the fiber is infinitesimal compared to any deviations of the flow from its local linear approximation.  If $\bp$ is the unit vector along the axis of symmetry of the fiber which is following the fluid, then it obeys the equation
\begin{equation}
\label{single}
\dot\bp = \bOmega\cdot\bp + \lambda\bGamma\cdot\bp - \lambda \bGamma:\bp\bp\bp .
\end{equation}
Here $\bOmega$ is the vorticity tensor, that is, the anti-symmetric part $\tfrac12(\bnabla\bu-(\bnabla\bu)^T)$ of the Jacobian  of the velocity field $\bnabla\bu = (\partial u_i/\partial x_j)_{1\le i,j\le3}$, and $\bGamma$ is the rate of strain tensor, that is, the symmetric part $\tfrac12(\bnabla\bu+(\bnabla\bu)^T)$ of the Jacobean of the velocity field.

This is commonly extended to a formula for the fiber orientation distribution $\psi = \psi(\bx,\bp,t)$, where $\bp$ is an element of the 2-dimensional sphere $S$.  Thus given a subset $E$ of $S$, the proportion of fibers whose direction is in $E$ is given by $\int_E \psi(\bx,\bp,t) \, d\bp$, where $d\bp$ represents the usual integration over $S$.  In particular, an isotropic distribution is represented by $\psi = 1/4\pi$.  Then Jeffery's equation for the fiber orientation distribution is
\begin{equation}
\label{dist}
\frac{D}{Dt}\psi = -\pnabla\cdot(\dot\bp \psi)
\end{equation}
where $D/Dt$ represents the material derivative $\partial/\partial t + \bu \cdot \bnabla$.  Equation~\eqref{dist} often includes a rotary diffusivity term to account for particle interaction, but that study is left for a related work in \cite{montgomery:10c}.

Since equation~\eqref{dist} is, in effect, a partial differential equation in 5-spatial dimensions, calculating this numerically can be rather daunting.  Hence \cite{hinch:76} suggested to recast the equation in terms of the higher order moment tensors.  For example, the second and fourth moment tensors are given by
\begin{equation}
\label{moment_tensor_definition}
A = \int_S \bp\bp\psi\,d\bp
, \qquad
\mathbb A = \int_S \bp\bp\bp\bp\psi\,d\bp.
\end{equation}
Then Jeffery's equation for the second order moment tensor is
\begin{equation}
\label{A}
\frac D{Dt}A = \bOmega\cdot A - A\cdot\bOmega + \lambda(\bGamma\cdot A+A\cdot \bGamma) - 2 \lambda \mathbb A:\bGamma.
\end{equation}
The difficulty with this equation is that it involves the fourth order moment tensor.  To circumvent this problem, various authors \cite[see e.g.,][]{hinch:76,altan:89,altan:90,verleye:93,cintra:95,verweyst:99,chung:02,han:02,jack:05a,jack:08b} have proposed \emph{closures}, that is, formulae to calculate the fourth order moment tensor $\mathbb A$ from the second order moment tensor $A$.  Many of the approximate closures involve coefficients obtained by fitting numerical data found by solving equation~\eqref{dist} using a finite element method \cite[see e.g.,][]{bay:91c}.

\cite{verleye:93} proposed what they called the \emph{natural closure}.  If there was an exact formula for the closure, then it must be the case that the formula is of the form
\begin{equation}
\label{nat-closure}
\mathbb A = \sum_{0 \le m \le n \le 2} \beta_{mn} \mathcal S(A^m \otimes A^n).
\end{equation}
Here $A^n$ denotes the matrix power, that is, the matrix $A$ multiplied by itself $n$ times, and $\mathcal S$ represents the symmetrization of a rank 4 tensor, that is
\begin{equation}
\mathcal S(\mathbb B)_{i_1i_2i_3i_4} = \tfrac1{24} \sum_\sigma \mathbb B_{i_{\sigma_1}i_{\sigma_2}i_{\sigma_3}i_{\sigma_4}},
\end{equation}
where the sum is over all permutations $\sigma=(\sigma_1,\sigma_2,\sigma_3,\sigma_4)$ of $(1,2,3,4)$. Furthermore, $\beta_{mn}$ are symmetric functions of the eigenvalues of $A$, or equivalently, functions of the three invariants of $A$, that is, the coefficients of the characteristic polynomial of $A$, which are the trace of $A$, the trace of the adjoint of $A$, and the determinant of $A$.  That $m$ and $n$ need only sum up to $2$ follows from the Cayley-Hamilton Theorem.

\cite{cintra:95} proposed what they called the \emph{orthotropic closure}.  If there was an exact formula for the closure, then it must be the case that if $A$ is diagonal, then $\mathbb A_{ijkl}$ must be zero if any index is repeated an odd number of times, and otherwise depend only on the diagonal entries of $A$ in such a manner that
\begin{subeqnarray}
\mathbb A_{1111}(A_{11},A_{22},A_{33}) &=& \mathbb A_{1111}(A_{11},A_{33},A_{22}), \\
\mathbb A_{1122}(A_{11},A_{22},A_{33}) &=& \mathbb A_{1122}(A_{22},A_{11},A_{33}), \\
\mathbb A_{2222}(A_{11},A_{22},A_{33}) &=& \mathbb A_{1111}(A_{22},A_{11},A_{33}), \\
\mathbb A_{3333}(A_{11},A_{22},A_{33}) &=& \mathbb A_{1111}(A_{33},A_{11},A_{22}), \\
\mathbb A_{1133}(A_{11},A_{22},A_{33}) &=& \mathbb A_{1122}(A_{11},A_{33},A_{22}), \\
\mathbb A_{2233}(A_{11},A_{22},A_{33}) &=& \mathbb A_{1122}(A_{22},A_{33},A_{11}).
\end{subeqnarray}
They set
\begin{subeqnarray}
\label{fgh}
\mathbb A_{1111}(A_{11},A_{22},A_{33}) &=& f(A_{11},A_{22},A_{33}), \\
\mathbb A_{2222}(A_{11},A_{22},A_{33}) &=& g(A_{11},A_{22},A_{33}), \\
\mathbb A_{3333}(A_{11},A_{22},A_{33}) &=& h(A_{11},A_{22},A_{33}),
\end{subeqnarray}
when $A_{11} \ge A_{22} \ge A_{33}$, and then searched for the three functions $f(x,y,z)$, $g(x,y,z)$ and $h(x,y,z)$.  They were then able to compute $\mathbb A_{1122}$, etc., using the relationship $\sum_i \mathbb A_{iijk} = A_{jk}$.

As noted in \cite{cintra:95}, since they come from identical assumptions, the natural closure and the orthotropic closure must be theoretically identical, in that the $\beta_{mn}$ must be uniquely determinable from the $\mathbb A_{iijj}$ and vice versa.  This can be seen explicitly by assuming that $A$ is diagonal in equation~\eqref{nat-closure}, and seeing that this creates six equations.  The transformation from the natural closure to the orthotropic closure is immediate.  The converse transformation is seen by solving the six equations for the six unknowns $\beta_{mn}$.  Calculation shows that the resulting six by six matrix is invertible if and only if the diagonal entries of $A$ are distinct.  If any of them are repeated, then the $\beta_{mn}$ can be obtained by a limiting process.  This suggests that while the natural and orthotropic closures are mathematically equivalent, and that numerically the orthotropic closure will be more stable when any of the eigenvalues of $A$ are repeated, and indeed numerical experiments seem to bare this out.

\cite{verleye:93} noted that indeed there is an exact closure, \emph{in the particular case that the fiber orientation distribution is at some time isotropic}.  Throughout this paper, this will be the standing assumption, unless otherwise stated.

Their exact closure followed from an exact solution to Jeffery's equation in this particular case provided by \cite{dinh:84} and \cite{lipscomb:88}.  Verleye and Dupret gave explicit formulae in the two dimensional case \cite[see also][who independently performed the same calculations]{altan:93}, but cited an `in preparation' article for the three dimensional case, merely noting that the calculations involved elliptic integrals.

Verweyst in his Ph.D.\ thesis \cite[][]{verweyst:98}, indicated similar exact calculations, for which he cited a `personal communication' from Wetzel.  Wetzel in his Ph.D.\ thesis \cite[][]{wetzel:99b} provides calculations for a different but related problem, and gives his solution in terms of elliptic integrals.  Therefore we are somewhat certain that he also obtained the formulae of Verleye and Dupret.

However, neither Dupret, Verleye, Verweyst, nor Wetzel seem to have published their calculations.  The traditional approach is to perform these sorts of calculations using spherical coordinates and the Legendre form of elliptic integrals.  However we prefer to use Cartesian coordinates on the sphere (which we think of as embedded in three dimensional Euclidean space) and the Carlson form of the elliptic integrals.  As such, it is quite possible that our formulae are more compact than the form they may have obtained.  Also, the approach taken here will easily generalize to other spatial dimensions, should the need arise.

Our analysis also yields what we shall call the \emph{Fast Exact Closure} (FEC), a rather fast method of solving Jeffery's equation.  This method calculates $\mathbb A$ by neither using approximate closures, nor computing elliptic integrals, but rather by computing an ancillary quantity $B$, which is easily computed via an ordinary differential equation.

We feel that the main contribution of this paper is to explicitly write out the exact formulae~\eqref{2nd slick} and~\eqref{4th slick}, which as far as we know have not been published elsewhere.  We also feel that a contribution of this paper is the Fast Exact Closure.  However this latter contribution is somewhat muted in that this is only valid for Jeffery's equation without any extra rotary diffusion terms.  We will amend this in another paper \cite[][]{montgomery:10c}.

Finally we note that some of the earlier publications asserted these formulae exist only in the case $\lambda=1$.  However following the work of \cite{lipscomb:88}, it was realized that this restriction is unnecessary.

\section{The Exact Formulae}
\label{Exact_Section}
\cite{bretherton:62} noticed that equation~\eqref{single} can be solved by $\bp = \bq/|\bq|$, where $\bq$ satisfies the equation
\begin{equation}
\label{qdot}
\dot\bq = (\bOmega+\lambda\bGamma)\cdot\bq .
\end{equation}
This can be seen by simple substitution.  From this it was found by \cite{dinh:84} (for the case $\lambda=1$) and \cite{lipscomb:88}, and later \cite{szeri:96}, that if the fiber orientation distribution is initially isotropic, that is, $\psi = 1/4\pi$ at $t=0$, then the solution to equation~\eqref{dist} is
\begin{equation}
\label{dist soln}
\psi(\bp) = \frac1{4\pi (B:\bp\bp)^{3/2}} .
\end{equation}
Here $B = C^T C$ where $C$ satisfies the equation
\begin{equation}
\label{C}
\frac{D}{D t} C = - C \cdot(\bOmega+\lambda \bGamma) ,\quad C = I \text{ at } t=0 ,
\end{equation}
or equivalently that $B$ satisfies
\begin{equation}
\label{B}
\frac{D}{D t} B = - B \cdot(\bOmega+\lambda \bGamma) - (-\bOmega+\lambda\bGamma)\cdot B ,\quad B = I \text{ at } t=0 .
\end{equation}
We note parenthetically that equation~\eqref{dist soln} generalizes as follows: if $\psi = \psi_0$ at $t=0$, \begin{equation}
\label{dist soln gen}
\psi(\bp) = \frac{\psi_0(C\cdot\bp/|C\cdot\bp|)}{(B:\bp\bp)^{3/2}} .
\end{equation}
For the purposes of the exact closure, the precise nature of $B$ is not required.  All that is necessary to know is that $B$ is a positive definite matrix with determinant one.

From the definition in equation~\eqref{moment_tensor_definition} along with equation~\eqref{dist soln}, the formula for the second order moment tensor may be expressed as
\begin{equation}
\label{2nd def}
A = \int_S \frac{\bp\bp}{4\pi (B:\bp\bp)^{3/2}} \, d\bp .
\end{equation}
Verleye and Dupret assert, without proof but somewhat reasonably, that this provides a one-one correspondence between positive definite matrices $B$ with determinant one, and positive definite matrices $A$ with trace one. By inverting the formula in equation~\eqref{2nd def}, one obtains a formula for $B$ in terms of $A$. We present the complete proof in the appendix showing that this expression provides a one-to-one and onto relationship between $A$ and $B$.

The integral in equation~\eqref{2nd def} is not sufficient for rapid computations, and we transfer the expression into the integral formula (see the derivation in the Appendix)
\begin{equation}
\label{2nd slick}
A = \tfrac12 \int_0^\infty \frac{(B+s I)^{-1} \, ds}{\sqrt{\text{det}(B+s I)}} .
\end{equation}
To perform this calculation numerically, diagonalize $B$ with an orthogonal matrix.  Then $A$ is transformed to a diagonal matrix via exactly the same similarity matrix, and equation~\eqref{2nd slick} becomes
\begin{subeqnarray}
\label{2nd long}
A_{11} &=& \tfrac12 \int_0^\infty \frac{ds}{(b_1+s)^{3/2}\sqrt{b_2+s}\sqrt{b_3+s}}, \\
A_{22} &=& \tfrac12 \int_0^\infty \frac{ds}{\sqrt{b_1+s}(b_2+s)^{3/2}\sqrt{b_3+s}}, \\
A_{33} &=& \tfrac12 \int_0^\infty \frac{ds}{\sqrt{b_1+s}\sqrt{b_2+s}(b_3+s)^{3/2}},
\end{subeqnarray}
where $b_1, b_2, b_3$ are the diagonal entries of $B$.  These integrals are one third the $r_D$ Carlson form of the elliptic integrals \cite[][]{carlson:95}.  The Carlson elliptic integrals have good numerical algorithms to compute them, for example, they are part of the suite of special functions provided by the GNU Scientific Library \cite[][]{gsl}, and are also described in the book \cite{press:03a}.    (We refer to the elliptic integrals as the Carlson form, because that is how it appears in \cite{press:03a}.  But such integrals appeared far earlier in the study of external ellipsoid harmonics, see for example \cite{oberbeck:1876}, \cite{hobson:31} or \cite{jeffery:23}.)

Combining equations~\eqref{moment_tensor_definition} and~\eqref{2nd def}, the fourth order moment tensor may be expressed as
\begin{equation}
\label{4th def}
\mathbb A = \int_S \frac{\bp\bp\bp\bp}{4\pi (B:\bp\bp)^{3/2}} \, d\bp,
\end{equation}
where we demonstrate in the appendix that this form may be transformed to
\begin{equation}
\label{4th slick}
\mathbb A = \tfrac34 \int_0^\infty \frac{s \, \mathcal S((B+s I)^{-1}\otimes(B+s I)^{-1}) \, ds}{\sqrt{\text{det}(B+s I)}},
\end{equation}
or in the special case that $B$ is diagonal (or equivalently when $A$ is diagonal) can be computed via the equations
\begin{subeqnarray}
\label{4th long}
\mathbb A_{1111} &=& \tfrac34 \int_0^\infty \frac{s\,ds}{(b_1+s)^{5/2}\sqrt{b_2+s}\sqrt{b_3+s}}, \\
\mathbb A_{2222} &=& \tfrac34 \int_0^\infty \frac{s\,ds}{\sqrt{b_1+s}(b_2+s)^{5/2}\sqrt{b_3+s}}, \\
\mathbb A_{3333} &=& \tfrac34 \int_0^\infty \frac{s\,ds}{\sqrt{b_1+s}\sqrt{b_2+s}(b_3+t)^{5/2}}, \\
\mathbb A_{1122} &=& \tfrac14 \int_0^\infty \frac{s\,ds}{(b_1+s)^{3/2}(b_2+s)^{3/2}\sqrt{b_3+s}}, \\
\mathbb A_{1133} &=& \tfrac14 \int_0^\infty \frac{s\,ds}{(b_1+s)^{3/2}\sqrt{b_2+s}(b_3+s)^{3/2}}, \\
\mathbb A_{2233} &=& \tfrac14 \int_0^\infty \frac{s\,ds}{\sqrt{b_1+s}(b_2+s)^{3/2}(b_3+s)^{3/2}},
\end{subeqnarray}
all the other coefficients of $\mathbb A$ being zero.  In the case that the eigenvalues of $B$ are distinct, these can be computed by the easily derived formulae
\begin{subeqnarray}
\label{4th quick}
\mathbb A_{1122} &=& \frac{b_1 A_{11} - b_2 A_{22}}{2(b_1-b_2)}, \\
\mathbb A_{1133} &=& \frac{b_1 A_{11} - b_3 A_{33}}{2(b_1-b_3)}, \\
\mathbb A_{2233} &=& \frac{b_2 A_{22} - b_3 A_{33}}{2(b_2-b_3)}, \\
\mathbb A_{1111} &=& A_{11} - \mathbb A_{1122} - \mathbb A_{1133}, \\
\mathbb A_{2222} &=& A_{22} - \mathbb A_{1122} - \mathbb A_{2233}, \\
\mathbb A_{3333} &=& A_{33} - \mathbb A_{1133} - \mathbb A_{2233}.
\end{subeqnarray}
If two of the eigenvalues are close to each other, for example $b_1=b_2 + \epsilon$, then equations~\eqref{4th quick} will give division by zero errors, or at least large numerical errors.  However, we see from equation~\eqref{4th long} that $A_{1111}+A_{2222}=6A_{1122} + O(\epsilon^2)$, and hence
\begin{subeqnarray}
\label{4th quick b1=b2+epsilon}
\mathbb A_{1133} &=& \frac{b_1 A_{11} - b_3 A_{33}}{2(b_1-b_3)}, \\
\mathbb A_{2233} &=& \frac{b_2 A_{22} - b_3 A_{33}}{2(b_2-b_3)} ,\\
\mathbb A_{1122} &=& \tfrac18(A_{11} + A_{22} - \mathbb A_{1133} - \mathbb A_{2233}) + O(\epsilon^2), \\
\mathbb A_{1111} &=& A_{11} - \mathbb A_{1122} - \mathbb A_{1133}, \\
\mathbb A_{2222} &=& A_{22} - \mathbb A_{1122} - \mathbb A_{2233}, \\
\mathbb A_{3333} &=& A_{33} - \mathbb A_{1133} - \mathbb A_{2233},
\end{subeqnarray}
where similar forms exist when $b_1=b_3 + \epsilon$ or $b_2=b_3 + \epsilon$. If all three of the eigenvalues are almost equal, that is $b_1=1+c_1$, $b_2=1+c_2$, $b_3=1+c_3$ with $c_1,c_2,c_3=O(\epsilon)$ (so necessarily $c_1+c_2+c_3 = O(\epsilon^2)$)
\begin{subeqnarray}
\label{4th quick b1=b2=b3+epsilon}
\mathbb A_{1111} &=& \tfrac15 - \tfrac3{14}c_1-\tfrac3{70}c_2-\tfrac3{70}c_3+O(\epsilon^2),\\
\mathbb A_{2222} &=& \tfrac15 - \tfrac3{70}c_1-\tfrac3{14}c_2-\tfrac3{70}c_3+O(\epsilon^2),\\
\mathbb A_{3333} &=& \tfrac15 - \tfrac3{70}c_1-\tfrac3{70}c_2-\tfrac3{14}c_3+O(\epsilon^2),\\
\mathbb A_{1122} &=& \tfrac1{15} - \tfrac3{70}c_1-\tfrac3{70}c_2-\tfrac1{70}c_3+O(\epsilon^2),\\
\mathbb A_{1133} &=& \tfrac1{15} - \tfrac3{70}c_1-\tfrac1{70}c_2-\tfrac3{70}c_3+O(\epsilon^2),\\
\mathbb A_{2233} &=& \tfrac1{15} - \tfrac1{70}c_1-\tfrac3{70}c_2-\tfrac3{70}c_3+O(\epsilon^2).
\end{subeqnarray}
For example, if you believe that your ODE solver is accurate to $n$ digits, it is appropriate to use \eqref{4th quick b1=b2+epsilon} or~\eqref{4th quick b1=b2=b3+epsilon} if any of the eigenvalues of $B$ have a relative difference bounded by about $10^{-n/3}$.  The resulting $\mathbb A$ should then be accurate to about $2n/3$ digits.  The hope is this loss of precision is unlikely to have a large effect on the final solution, because the event that $B$ has eigenvalues close to each other should happen somewhat rarely, and these events will largely be integrated out by equation~\eqref{B}.

The exact closure can be summarized by first diagonalizing $A$, then computing $B$ by inverting the formulae~\eqref{2nd long}, then computing $\mathbb A$ using \eqref{4th long}, and then using the inverse of the similarity matrix from the diagonalization of $A$ to transform $\mathbb A$.  To compute the formulae~\eqref{4th long}, use \eqref{4th quick}, \eqref{4th quick b1=b2+epsilon}, or~\eqref{4th quick b1=b2=b3+epsilon}, as appropriate.

These numerical calculations will necessarily take some time to perform (particularly the inversion process which requires a multidimensional root finding). \cite{verweyst:98} described a different process where these formulae were used to generate values which were then fitted to polynomial approximations to the functions $f$, $g$ and $h$ as in equation~\eqref{fgh}.

\section{The Fast Exact Closure (FEC)}
\label{FEC_Section}
Our analysis yields another method for solving Jeffery's equation, with the proviso that the probability distribution was isotropic at some time in its history where the goal is to solve equation~\eqref{A}.  The difficulty of finding $B$ can be solved by simultaneously solving equation~\eqref{B} along with equation~\eqref{A}.  During this process, $\mathbb A$ can be calculated using \eqref{4th quick}-\eqref{4th quick b1=b2=b3+epsilon}.  In this way, an exact solution is obtained, without the need to compute the elliptic integrals.

The only difficulty will be with the initial data.  If it is isotropic, one simply sets $A=\tfrac13I$ and $B=I$ at $t=0$. Conversely, if the initial data is not isotropic, then the initial data for $B$ can be calculated by inverting equation~\eqref{2nd slick}.  This can be performed by diagonalizing $A$, computing $B$ by inverting equation~\eqref{2nd long}, and then reversing the change of basis that was used to diagonalize $A$.  While this calculation could conceivably be quite slow, as the inversion of equation~\eqref{2nd long} will require a three-dimensional root finding method as well as computing the elliptic integrals, this needs to be performed only for the initial step.

This method should run at about the same speed as computing the orthotropic closure algorithms of \cite{cintra:95}, or \cite{verweyst:99}.  The work load of this algorithm involves solving an ordinary differential equation in about eleven variables, where the main computational effort will be the diagonalization of the  matrices, and performing the basis change for the rank two and rank four tensors.  (Note that $A$ and $B$ will be simultaneously diagonalizable, that is, the orthogonal matrix used to diagonalize one of the second-order tensors can and should be used to diagonalize the other.)

A major problem with this fast exact closure is that it only works in the case when there is no rotary diffusion term.  But the pure Jeffery's equation without a rotary diffusion term is only applicable to a single fiber in a fluid, and so is unrealistic.  Resolving this issue is a major extension of this paper, and we intend to address this in a future paper \cite[][]{montgomery:10c}.

\section{Behavior of the Fast Exact Closure (FEC)}
\label{Results_Section}
Direct solutions of the tensor $C$ of equation~\eqref{C} will serve as the basis upon which to compare the effectiveness of the FEC. The transient solution of $C$ in equation~\eqref{C}, assuming constant velocity gradients, is given by
\begin{equation}
\label{C_soln}
C=\exp(-t(\bOmega+\lambda\bGamma)) ,
\end{equation}
where $\exp$ denotes the matrix exponential. The solution for $B$ readily follows from the definition $B=C^TC$ from the solution of $C$ in equation~\eqref{C}, and the second-order orientation tensor $A$ is obtained from equation~\eqref{2nd def}. This solution approach, which we term the analytic solution, is only effective for flows originating from an isotropic initial condition with constant velocity gradients, but will serve as the basis upon which to compare results from various closures which numerically solve the equation of motion for the orientation tensor $A$ of equation~\eqref{A}.

To demonstrate the effectiveness of the FEC, solutions are compared to results from several popular closures, the quadratic \cite[][]{doi:81}, the hybrid \cite[][]{advani:87a} and the ORT \cite[][]{verweyst:98,verweyst:02}, and all results will be compared to the analytic solutions from equation~\eqref{C_soln}. The three comparison closures are selected based on their common employment in industrial simulations of fiber orientation (e.g., the quadratic and the hybrid) and their known accuracy (e.g., the ORT).  It is expected the ORT closure's accuracy will be quite high considering it is based off of similar `exact' calculations \cite[][]{verweyst:98} to the FEC. The ORT is known to give similar results to the IBOF closure of \cite{chung:02} which is an extension of the natural closure discussed by \cite{verleye:93}. Solutions for the existing closures (quadratic, hybrid, or ORT) are performed by solving equation~\eqref{A} and using the respective closure to approximate $\mathbb A$ in terms of $A$.

The simple shear flow considered (${\bf v}=\left[Gx_3,0,0\right]^T$ where $G$ is the scalar magnitude of the rate of deformation $G=||\bGamma||$) begins from an initial isotropic orientation (i.e., $\psi=\frac{1}{4\pi}$). The primary reason for choosing a pure shearing flow is twofold. The first, and primary, reason is due to the prevalence of shearing dominated flows in industrial injection molding processes. An alternative, yet more constraining, motivation for choosing a pure shearing flow is to demonstrate the FEC's ability to capture the correct period for the Jeffery orbit. It is this oscillatory nature of the orientation in pure Jeffery's motion that causes significant numerical difficulties for classical approaches that rely on alternative methods such as finite differences \cite[][]{bay:91a,bay:91c} or the Spherical Harmonic Expansion approach \cite[see e.g.,][]{bird:87a,montgomery:10a}. It is worthwhile to note that the interior of the eigenspace triangle containing all possible orientations \cite[see e.g.,][for a full discussion]{cintra:95} with an initial isotropic orientation state can only be populated by shearing flows, whereas orientations occurring on the boundaries are caused by pure elongational flows and will have two repeated eigenvalues of $A$ and $B$.

\begin{figure}
\centering
\begin{tabular}{ccc}
\includegraphics[scale=0.28]{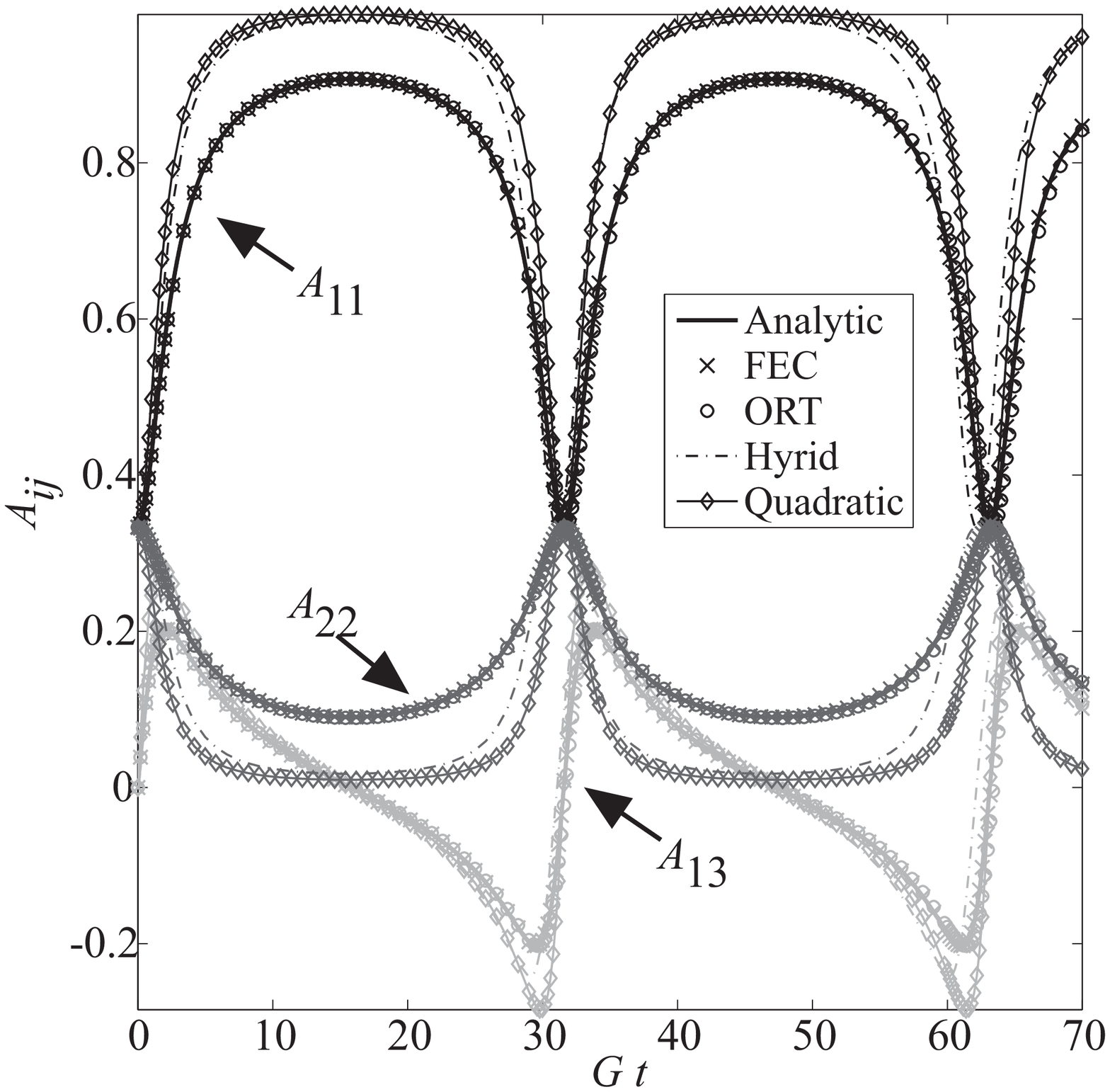}  &  \begin{tabular}[b]{c}$\longrightarrow$\\\\\\\\\\\\\\\end{tabular}  &  \includegraphics[scale=0.28]{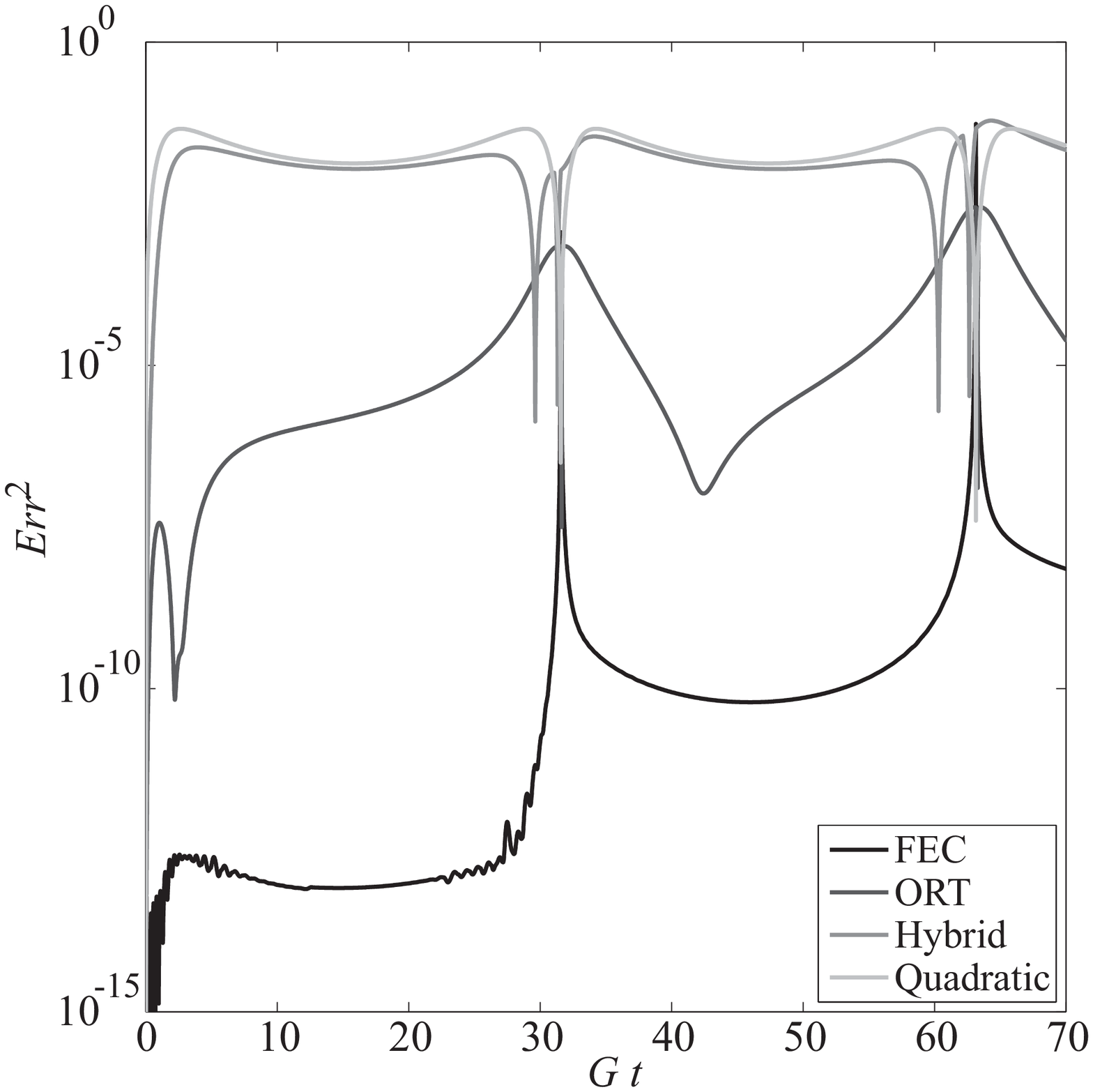}
\end{tabular}
\caption{Plot of $A_{11}$ for Simple Shear flow with $\lambda=0.98$ from various closures along with the associated error $Err$.}
\label{fig:SS_lam_98}
\end{figure}

Solutions of pure shear flow are run for the orientation parameter $\lambda=0.98$, which corresponds to a fiber aspect ratio of $a_r=L/d\simeq 9.95$. The results for the second-order orientation tensor $A$ from each solution approach is given in Figure~\ref{fig:SS_lam_98} and compared to direct computations of the analytic solution of equation~\eqref{C_soln} for the tensor $C$ and then using the elliptic integral relationship of equation~\eqref{2nd long} to obtain $A$. As observed in the figure, the periodic nature of the orientation solution of $A_{ij}$ is captured reasonably will by each of the solution approaches. An ellipsoid experiencing Jeffery motion in a pure shearing flow will rotate with a period, $\tau$, of \cite[][]{jeffery:23}
\begin{equation}
\label{Jeff_Period}
\tau=\frac{2\pi}{||\bGamma||\sqrt{\left(1-\lambda^2\right)}} ,
\end{equation}
where the solution of equation~\eqref{Jeff_Period} may be obtained by analyzing the eigenvalues of $\bOmega+\lambda\bGamma$ from equation~\eqref{qdot}. In our example $\lambda=0.98$ and $||\bGamma||=G$ and thus $\tau=31.574/G$. This period can be observed by noting a particular orientation condition, and tabulating the time it takes to return to the same orientation state. Since there are numerical errors introduced by the numerical solution of a system of ODEs with a periodic solution, returning to the numerically identical orientation state may never occur.  Thus, we compare the initial isotropic orientation to that of the orientation occurring near where the three eigenvalues are nearest to each other and can approximate the period of the numerical solution for pure show flow through the expression
\begin{equation}
\label{tau_0_98}
\tau_{\mbox{\tiny Closure}} = t \text{ such that }|a_1-a_3|\text{ is minimized over }Gt\in[30,33] ,
\end{equation}
where $a_1$, $a_2$ and $a_3$ are the eigenvalues of the second-order orientation tensor $A$ in descending order, $a_1\geq a_2\geq a_3\geq 0$. Therefore, the orientation from the numerical solutions employing closures will be closest to isotropic ($a_1=a_2=a_3=1/3$) when $a_1\simeq a_3$. The FEC, ORT, Hybrid and Quadratic closures have observed periods of, respectively, $G\tau_{\mbox{\tiny FEC}}=31.572$, $G\tau_{\mbox{\tiny ORT}}=31.708$, $G\tau_{\mbox{\tiny Hybrid}} = 31.081$, and $G\tau_{\mbox{\tiny Quadratic}}=31.575$. Each of the closures are reasonably accurate, whereas the FEC and Quadratic closures are the most effective in representing the actual period. But by observing the orientation path taken by each of the closures, the quadratic and the hybrid are clearly ineffective at predicting the orientation state during most of the flow history. The FEC and ORT closures follow the analytic solution quite well. To aid in comparing between closures, the error is plotted in Figure~\ref{fig:SS_lam_98} where the error is defined as the difference in the eigenvalues from the analytic solution and the numeric solution from the closures as
\begin{equation}
\label{err}
Err^2=\sum_{i=1}^2\left(a_i^{\mbox{\tiny Analytic}}-a_i^{\mbox{\tiny Closure}}\right)^2 .
\end{equation}
The value of $Err$ shows the effectiveness of a closure in conjunction with the solution of the ODEs in equations~\eqref{A} and~\eqref{B} and where it has the greatest deviation in the orientation solution.  It is interesting to note that the FEC performs quite well throughout much of the flow history.  This characteristic is true, except when the orientation approaches isotropic. This is not unexpected due to the nature of the closure since this is where the eigenvalues of $A$ all approach the same value of $1/3$ and the approximation of equation~\eqref{4th quick b1=b2=b3+epsilon} must be accepted due to the aforementioned numerical limitations of repeated eigenvalues. It is worthwhile to note that both the ORT and the FEC yield solutions that are a considerable improvement over the commonly employed Quadratic and Hybrid closures, which for the ORT is not surprising and this characteristic has been widely discussed in the literature. It is not intended that the reader should infer that the FEC is proposed as an improvement over the quite sufficient ORT closure, where the accuracy to the third decimal place observed by the ORT closure is considered quite acceptable in injection molding predictions (see e.g., \cite{chung:02}), but the FEC serves as a starting point for the development of a new class of closures for fiber systems with colliding particles which add a diffusion term to equation~\eqref{dist} \cite[see e.g.,][for a listing of existing diffusion models]{montgomery:10a}. For the industrially applicable case of concentrated suspensions of fibers, extensions to the FEC have been performed by the authors in \cite{montgomery:10c} for Jeffery's equation with rotary diffusion.

The authors have performed similar studies for flows beyond pure shear, and the results and conclusions for virtually all flows are identical to those from simple shear. It is expected that flows with repeated eigenvalues (i.e., orientations occurring on the boundaries of the eigenspace triangle containing all physically possible orientations discussed in \cite{cintra:95}) would cause problems due to the numerical approximation that must be made in equations~\eqref{4th quick b1=b2+epsilon} or~\eqref{4th quick b1=b2=b3+epsilon}. But for flows that have some non-zero shear coupled with elongation, as would be expected in virtually all industrial flows, the FEC and the ORT closure both perform quite well and the aforementioned conclusions for the accuracy and stability for pure shear were equivalent for all flows investigated with mixed shearing and uniaxial elongation. The primary difference between the pure shear flow and the shear-elongation flows is the absence of any periodic behavior due to the elongation, and thus the numerical issues in the ODE solution are greatly diminished as $t\rightarrow\infty$. It is noted that coupled flows with shearing and biaxial elongation were also performed, and for cases when biaxial elongation dominated the flow, numerical instabilities were observed in the FEC since $a_1\simeq a_2$ remains true throughout the entire flow history.  This case is worthwhile to recognize, but is not of a significant concern as this orientation state will rarely, if ever, occur in actual industrial processes.

\section{Conclusion}
\label{Conclusion_Section}
In this paper, we have presented an exact closure, which was at least in part, understood by previous authors.  We have found that an analysis of the formulae gives rise to the Fast Exact Closure (FEC), which exactly solves Jeffery's equation in a fast manner for any orientation that at one point in time was isotropic. This work demonstrates that the observations by previous authors were correct that solutions of the Jeffery equation must exist.  The FEC closure presents a systematic approach to solve the Jeffery's equation while simultaneously solving a related differential equation, thus avoiding both the cumbersome elliptic integral evaluation and the necessity for a closure of a higher order moment tensor. It is noted that industrial simulations often add in a term like $C_I \|\bGamma\|(2I-6A)$ to equation~\eqref{A}, and then presume that the closure that creates $\mathbb A$ from $A$ is still valid.  However, to add diffusion terms to the FEC method requires adding a similar term to equation~\eqref{B}. For the work which explores approximations to versions of Jeffery's equation with diffusion terms added, for example, the Folgar-Tucker equation \cite{folgar:84}, the interested reader is directed toward \cite{montgomery:10c}.

\acknowledgments
The authors gratefully acknowledge support from N.S.F. via grant C.M.M.I. 0727399.

\bibliography{info_10}
\bibliographystyle{jfm}

\section{Appendix: Justification of the Formulae}
\label{Proof_Section}

First we show that the initially isotropic orientation distribution function from equations~\eqref{dist soln} and~\eqref{dist soln gen} are solutions to equation~\eqref{dist}.  Even though this is not a new result, we present our proof, firstly to keep the presentation self-contained, and secondly because we later use the intermediate equation~\eqref{int 2}.  However the authors do feel that the proof in \cite{szeri:96} is superior.

If $E$ is a subset of $S$, then the proportion of fibers whose direction is contained in $E$ is
\begin{equation}
\label{int 1}
\int_E \psi(\bp) \, d\bp = \frac{4}{\pi ^{1/2}} \int_{r=0}^\infty \int_{\bp \in E} \psi(\bp) r^2 e^{-r^2} \, d\bp \, dr
= \frac{1}{\pi^{3/2}} \int_{\bq \in \bar E} \psi\left(\frac{\bq}{|\bq|}\right) e^{-|\bq|^2} \, d\bq .
\end{equation}
Here $\bar E$ denotes the set of elements $\bq \in \mathbb R^3-\{0\}$ such that $\bq/|\bq| \in E$.  The first equality follows because $\int_0^\infty r^2 e^{-r^2} \, dr = \pi^{1/2}/4$.  The second equality is just the conversion from spherical coordinates to Cartesian coordinates, noting that $4\pi r^2 dr d\bp = d\bq$.

Note that equation~\eqref{qdot} is solved by setting $\bq = C^{-1}\cdot\bp(0)$ where $C$ satisfies equation~\eqref{C}, because $C^{-1}$ satisfies the equation
\begin{equation}
\label{Cinv}
\frac{D}{D t} C^{-1} = (\bOmega+\lambda \bGamma) \cdot C^{-1} ,\quad C^{-1} = I \text{ at } t=0.
\end{equation}
Thus a fiber which at time $t$ is in a direction $\bp$ will have come from a fiber which at time $t=0$ was in the direction $\bp_0 = C\cdot\bp/|C\cdot\bp|$.  The set $E$ will have come from the set $E_0 = \{C\cdot\bp/|C\cdot\bp|:\bp\in E\}$, and the set $\bar E$ will have come from the set $\bar E_0 = C\cdot \bar E:=\{C\cdot\bq:\bq\in \bar E\}$.  Thus
\begin{equation}
\int_{E} \psi(\bp) \, d\bp = \int_{E_0} \psi_0(\bp) \, d\bp .
\end{equation}
Rewriting equation~\eqref{int 1} for the time $t=0$, we obtain
\begin{equation}
\int_{E_0} \psi_0(\bp) \, d\bp = \pi^{-3/2} \int_{\bq \in \bar E_0} \psi_0\left(\frac{\bq}{|\bq|}\right) e^{-|\bq|^2} \, d\bq .
\end{equation}
Performing the change of variables $\bq \mapsto C\cdot\bq$, and noting that $d\bq = d(C\cdot\bq)$ (because the determinant of $C$ is one), we obtain
\begin{equation}
\begin{split}
\int_{E_0} \psi_0(\bp) \, d\bp &= \pi^{-3/2} \int_{\bq \in \bar E} \psi_0\left(\frac{C\cdot\bq}{|C\cdot\bq|}\right) e^{-|C\cdot\bq|^2} \, d\bq \\
&= \pi^{-3/2} \int_{\bp \in E} \int_{r=0}^\infty \psi_0\left(\frac{C\cdot\bp}{|C\cdot\bp|}\right) r^2 e^{-|C\cdot\bp|^2 r^2}\, dr d\bp ,
\end{split}
\end{equation}
where the last equality follows by converting Cartesian coordinates to spherical coordinates.  Combining this with equation~\eqref{int 1}, and noting that these equations work for any subset $E$ of $S$, we obtain
\begin{equation}
\label{int 2}
\psi(\bp) = 4\pi^{-1/2} \psi_0\left(\frac{C\cdot\bp}{|C\cdot\bp|}\right) \int_{r=0}^\infty r^2 e^{-|C\cdot\bp|^2 r^2}\, dr = \frac{\psi_0(C\cdot\bp/|C\cdot\bp|)}{|C\cdot\bp|^3}
\end{equation}
as required.  Equation~\eqref{dist soln} is an immediate corollary of equation~\eqref{dist soln gen}.

Next, we show that the transformation of the second-order orientation tensor of equations~\eqref{2nd slick} and~\eqref{2nd long} from equation~\eqref{2nd def} is valid.
Note that if $B$ is diagonal with diagonal entries $b_1,b_2,b_3$, then from equation~\eqref{int 2}
\begin{equation}
\psi(\bp) = \frac1{\pi^{3/2}}\int_0^\infty r^2 \exp(-r^2(b_1 p_1^2 + b_2 p_2^2 + b_3 p_3^2)) \, dr .
\end{equation}
By symmetry we can see that the off diagonal entries of $A$ are zero, and that the diagonal entries are computed, for example, by
\begin{equation}
\begin{split}
A_{11}
&= \frac1{\pi^{3/2}} \int_S \int_{r=0}^\infty r^2 p_1^2 \exp(-r^2(b_1 p_1^2 + b_2 p_2^2 + b_3 p_3^2)) \, dr d\bp \\
&= \frac1{\pi^{3/2}} \int_{\mathbb R^3} \frac{q_1^2}{|\bq|^2} \exp(-(b_1 q_1^2 + b_2 q_2^2 + b_3 q_3^2)) \, d\bq .
\end{split}
\end{equation}
Now, if we set
\begin{equation}
f(s) = \frac1{\pi^{3/2}} \int_{\mathbb R^3} \frac{q_1^2}{|\bq|^2} \exp(-((b_1+s)q_1^2 + (b_2+s)q_2^2 + (b_3+s)q_3^2)) \, d\bq ,
\end{equation}
then we see that
\begin{equation}
\begin{split}
f'(s) &= -\frac1{\pi^{3/2}} \int_{\mathbb R^3} q_1^2 \exp(-((b_1+s)q_1^2 + (b_2+s)q_2^2 + (b_3+s)q_3^2)) \, d\bq \\
&= -\frac1{\pi^{3/2}} \int_{-\infty}^\infty q_1^2 e^{-(b_1+s)q_1^2} \, dq_1
   \int_{-\infty}^\infty e^{-(b_2+s)q_2^2} \, dq_2
   \int_{-\infty}^\infty e^{-(b_3+s)q_3^2} \, dq_3 \\
&= -\frac1{2(b_1+s)^{3/2}\sqrt{b_2+s}\sqrt{b_3+s}} .
\end{split}
\end{equation}
Since it is evident that $f(s)\to0$ as $s\to\infty$, we obtain
\begin{equation}
A_{11} = \tfrac12 \int_0^\infty \frac{ds}{(b_1+s)^{3/2}\sqrt{b_2+s}\sqrt{b_3+s}} .
\end{equation}
To show equation~\eqref{2nd slick}, it is only necessary to consider the case when $B$ is diagonal, and in this case the formula follows immediately from \eqref{2nd long}.

Next, we show that the transformation of the fourth-order orientation tensor of equations~\eqref{4th slick} and~\eqref{4th long} from equation~\eqref{4th def} is valid.
We illustrate the argument with the example
\begin{equation}
\begin{split}
\mathbb A_{1111}
&= \frac1{\pi^{3/2}} \int_S \int_{r=0}^\infty r^2 p_1^4 \exp(-r^2(b_1 p_1^2 + b_2 p_2^2 + b_3 p_3^2)) \, dr d\bp \\
&= \frac1{\pi^{3/2}} \int_{\mathbb R^3} \frac{q_1^4}{|\bq|^4} \exp(-(b_1 q_1^2 + b_2 q_2^2 + b_3 q_3^2)) \, d\bq .
\end{split}
\end{equation}
We set
\begin{equation}
g(s) = \frac1{\pi^{3/2}} \int_{\mathbb R^3} \frac{q_1^4}{|\bq|^4} \exp(-((b_1+s)q_1^2 + (b_2+s)q_2^2 + (b_3+s)q_3^2)) \, d\bq ,
\end{equation}
and we see that
\begin{equation}
\begin{split}
g''(s) &= \frac1{\pi^{3/2}} \int_{\mathbb R^3} q_1^4 \exp(-((b_1+s)q_1^2 + (b_2+s)q_2^2 + (b_3+s)q_3^2)) \, d\bq \\
&= \frac1{\pi^{3/2}} \int_{-\infty}^\infty q_1^4 e^{-(b_1+s)q_1^2} \, dq_1
   \int_{-\infty}^\infty e^{-(b_2+s)q_2^2} \, dq_2
   \int_{-\infty}^\infty e^{-(b_3+s)q_3^2} \, dq_3 \\
&= \frac3{4(b_1+s)^{5/2}\sqrt{b_2+s}\sqrt{b_3+s}}
\end{split}
\end{equation}
Since
\begin{equation}
g(s) = \int g'(s) \, ds = s g'(s) - \int s g''(s) \, ds ,
\end{equation}
it follows that
\begin{equation}
\mathbb A_{1111} = \tfrac34 \int_0^\infty \frac{s\,ds}{(b_1+s)^{5/2}\sqrt{b_2+s}\sqrt{b_3+s}} .
\end{equation}
To show equation~\eqref{4th slick}, again it is only necessary to prove this in the case when $B$ is diagonal.  The following identities are useful.  If $U$ and $V$ are diagonal matrices, then
\begin{equation}
\begin{split}
\mathcal S(UV)_{iiii} &= U_{ii}V_{ii} \\
\mathcal S(UV)_{iijj} &= \tfrac16(U_{ii}V_{jj} + U_{jj}V_{ii}) \quad \text{ if $i \ne j$} \\
\mathcal S(UV)_{iijk} &= 0 \quad \text{ if $i \ne j$ and $j\ne k$} ,
\end{split}
\end{equation}
where no summation is implied for repeated indices.

\begin{figure}
\centering
\begin{tabular}{ccc}
\includegraphics[scale=0.5]{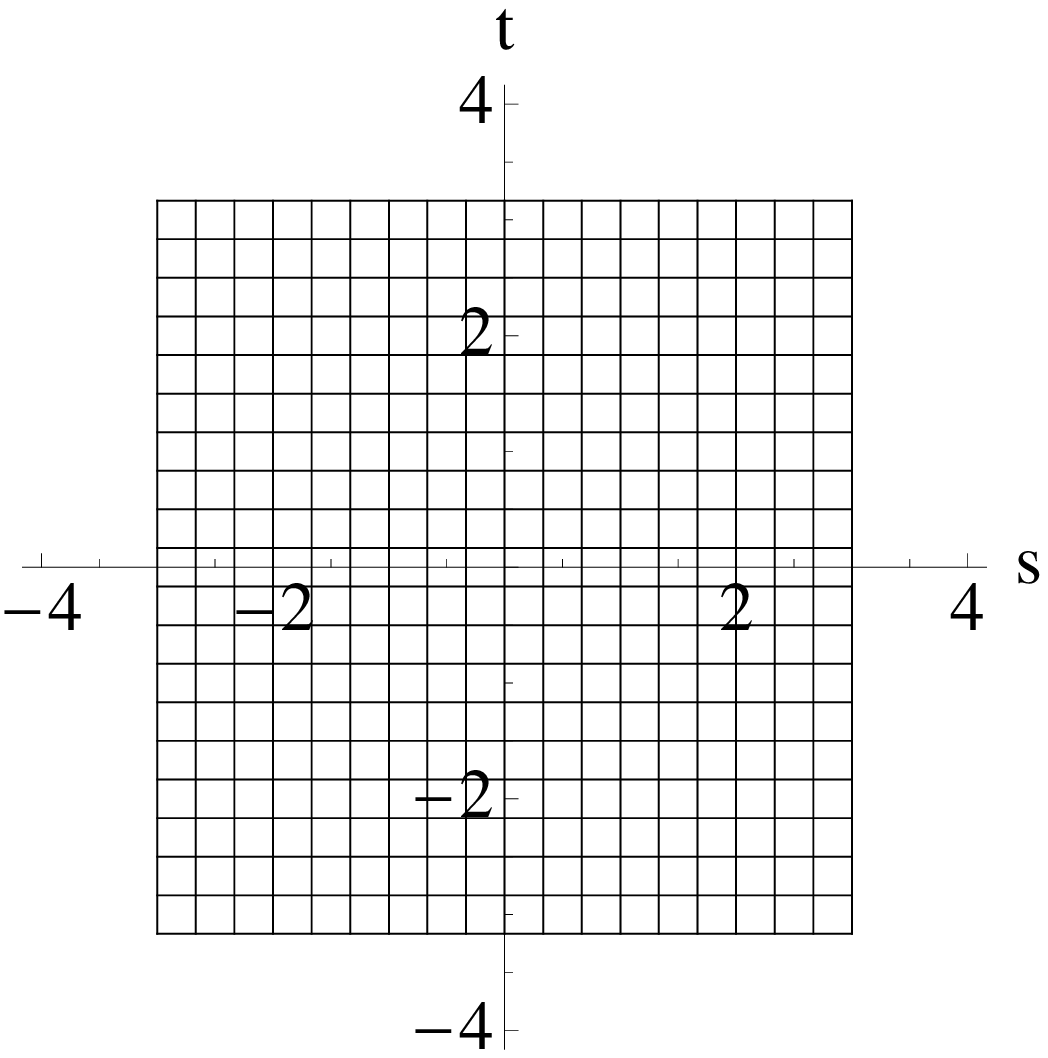}  &  \begin{tabular}[b]{c}$\longrightarrow$\\\\\\\\\\\\\\\end{tabular}  &  \includegraphics[scale=0.5]{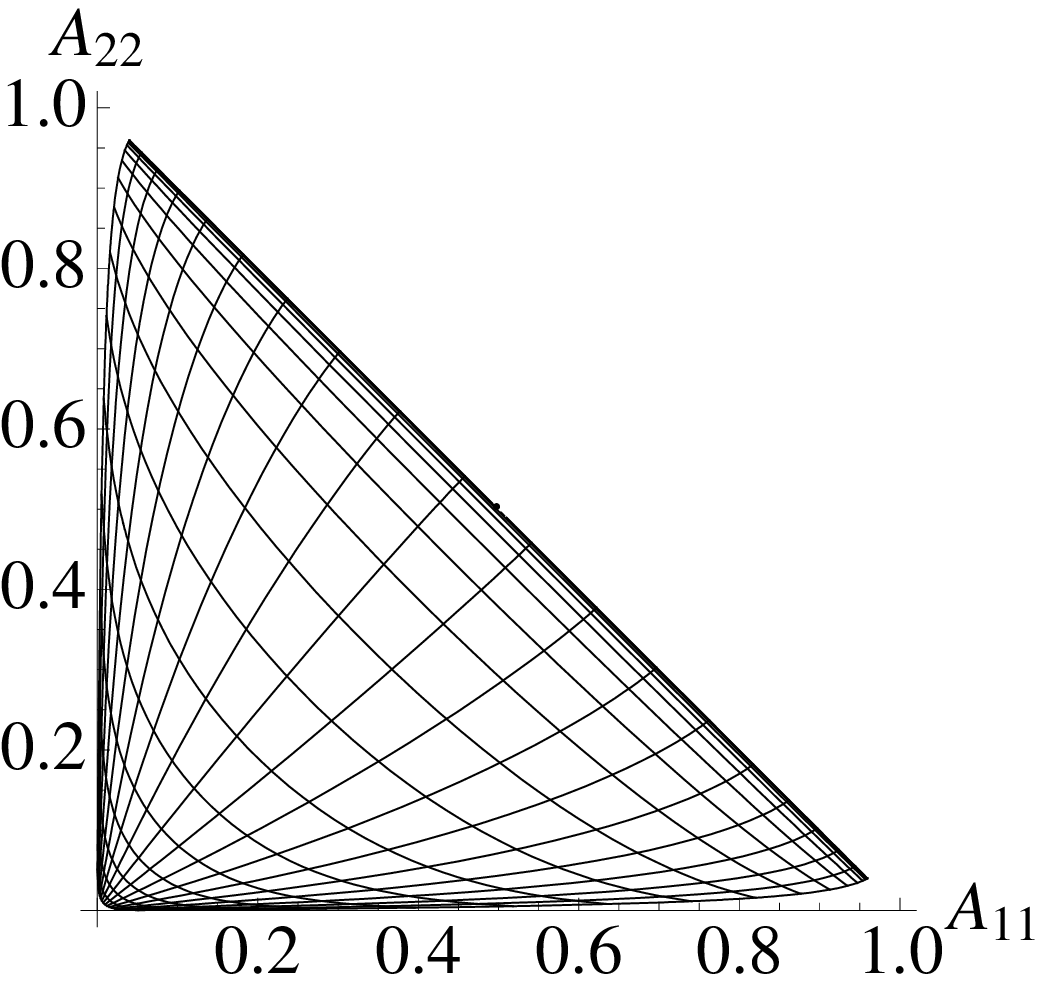}
\end{tabular}
\caption{Diagram illustrating how $\gamma$ maps a grid in the $s$-$t$ plane to the $A_{11}$-$A_{22}$ plane.}
\label{fig:plot}
\end{figure}

Next, we show that the mapping from $B$ to $A$ defined by equation~\eqref{2nd slick} is one-one and onto.
This is equivalent to showing that the map from $\{(b_1,b_2,b_3):b_1,b_2,b_3>0,\,b_1b_2b_3=1\}$ to $\{(A_{11},A_{22},A_{33}):A_{11},A_{22},A_{33}>0,\,A_{11}+A_{22}+A_{33}=1\}$ defined by equation~\eqref{2nd long} is one-one and onto.  To this end, consider the map $\gamma$ that takes $(s,t)$ to the values of $(A_{11},A_{22})$ when $(b_1,b_2,b_3) = (e^{s+t},e^{s-t},e^{-2s})$.  In Figure~\ref{fig:plot}, we show how $\gamma$ maps a grid in the $s$-$t$ plane to the $A_{11}$-$A_{22}$ plane, providing a `picture proof' that the map is one-one and onto.

To show that the map is one-one, it is sufficient to show that $\gamma(s,t)$ gives a family of curves with negative slope if $s$ is fixed, and $t$ is allowed to range freely, and that $\gamma(s,t)$ gives a family of curves with positive slope if the roles of $s$ and $t$ are reversed.

For example, let us prove that $\partial A_{11}/\partial t < 0$.  We have that $\partial B/\partial t = G\cdot B$ with $G=\left[\begin{smallmatrix}1&0&0\\0&-1&0\\0&0&0\end{smallmatrix}\right]$, that is, $B$ satisfies equation~\eqref{B} with $\bGamma=-G$, $\bOmega=0$, $\lambda=1$ and $\bu=0$.  (Since $B$ is diagonal, we have $B\cdot G=G\cdot B$.)  Hence $A$ satisfies equation~\eqref{A} with the same parameters, and so $\partial A_{11}/\partial t = -A_{11}+\mathbb A_{1111}-\mathbb A_{1122} = -2\mathbb A_{1122}-\mathbb A_{1133} < 0$.  Similarly, $\partial A_{22}/\partial t > 0$, and hence $\gamma(s,t)$ gives a family of curves with negative slope if $s$ is fixed and $t$ is allowed to range freely.

To show that the map is onto, it is sufficient to see that the map $\gamma$ `stakes out' the appropriate region, which follows because $\gamma(s,t)\to(0,0)$ as $s\to\infty$, $\gamma(s,t)$ converges to a point on the line $A_{11}+A_{22}=1$ as $s\to-\infty$, $A_{11}\to0$ as $t\to\infty$, and $A_{22}\to0$ as $t\to-\infty$.

For example, as $s\to-\infty$, we see from equation~\eqref{dist soln} that the fiber orientation distribution concentrates on the equator $p_3=0$.  Hence $A_{33} \to 0$, and so $A_{11}+A_{22} = 1-A_{33} \to 1$.  Since $A_{11}$ and $A_{22}$ increase as $s$ decreases, they must also individually converge to limits.

\label{lastpage}

\end{document}